\documentclass[aps,prd,10pt,floats,floatfix,showpacs,amssymb,twocolumn,superscriptaddress,nofootinbib]{revtex4-2}

\usepackage{graphicx}
\usepackage{dcolumn}
\usepackage{bm}
\usepackage{amsmath}
\usepackage[colorlinks=true,linkcolor=blue,citecolor=blue,urlcolor=blue]{hyperref}
\usepackage{dsfont}
\usepackage{orcidlink}
\usepackage{booktabs}
\usepackage{adjustbox}
\usepackage{placeins}

\begin{document}

\title{\texorpdfstring{Primordial Magnetic Fields at Cosmic Dawn: \\ 21-cm Forecasts with HERA and SKA}{Primordial Magnetic Fields at Cosmic Dawn: 21-cm Forecasts with HERA and SKA}}
\date{\today}

\author{Keduse Worku\orcidlink{0000-0002-2289-5541}}
\affiliation{William H.\ Miller III Department of Physics \& Astronomy, Johns Hopkins University, 3400 N.\ Charles St., Baltimore, MD 21218, USA}

\author{Hector Afonso G. Cruz\orcidlink{0000-0002-1775-3602}}       
\affiliation{Center for Cosmology and Particle Physics, Department of Physics,
New York University, New York, New York 10003, USA}

\author{Marc Kamionkowski\orcidlink{0000-0001-7018-2055}}       
\affiliation{William H.\ Miller III Department of Physics \& Astronomy, Johns Hopkins University, 3400 N.\ Charles St., Baltimore, MD 21218, USA}

\begin{abstract}
Primordial magnetic fields (PMFs) can enhance the abundance of low-mass halos during Cosmic Dawn by sourcing additional small-scale matter fluctuations. This enhanced small-scale power can accelerate early galaxy formation, shifting the timing of Lyman-$\alpha$ coupling, X-ray heating, and reionization toward earlier times and imprinting correlated signatures on the global and fluctuating 21-cm signals. We extend the fast analytic framework {\tt\string zeus21} to include a physically motivated PMF contribution to the linear matter power spectrum, including radiative damping before recombination and magnetic-pressure suppression below the magnetic Jeans scale. The implementation preserves the speed and modularity of {\tt\string zeus21}, enabling efficient exploration of PMF parameter space. For $n_B=-2.9$, we quantify the impact of PMFs on early structure formation and 21-cm observables across a range of fiducial magnetic amplitudes, and forecast detectability with \textit{HERA} and \textit{SKA}. Combining 21-cm forecasts with external CMB priors, we find that upcoming experiments can probe PMFs through their impact on small-scale structure, providing constraints complementary to existing cosmological probes.
\end{abstract}

\maketitle

\section{Introduction}
The redshifted 21-cm signal from neutral hydrogen provides a unique probe of early structure formation during Cosmic Dawn and the Epoch of Reionization (EoR). During this era, the first stars and galaxies transformed the thermal and ionization state of the intergalactic medium (IGM)~\citep{Loeb:2003ya, Barkana:2004zy, Furlanetto:2006jb, Pritchard:2011xb, Koopmans:2015sua}. Unlike traditional galaxy surveys, which provide discrete Poisson tracers of the underlying matter density field, the 21-cm signal traces neutral hydrogen throughout the IGM as a continuous field. Because this emission arises before galaxies become abundant and bright, it enables direct probes of structure formation and thermal evolution at redshifts and on physical scales inaccessible to galaxy observations alone~\citep{Furlanetto:2006jb, Pritchard:2011xb}. Current and upcoming 21-cm experiments, including the Hydrogen Epoch of Reionization Array (HERA) and the Square Kilometre Array (SKA), are expected to deliver constraints on Cosmic Dawn and EoR. These measurements will enable tighter inference on early star formation, X-ray heating, and the topology of ionized regions~\citep{Fragos:2012vf, Fragos:2013bfa, Mellema:2012ht, Parsons_2012, Koopmans:2015sua, DeBoer:2016tnn}. The 21-cm signal is also sensitive to physics beyond the concordance $\Lambda$CDM model of cosmology, particularly to scenarios that enhance small-scale density fluctuations before the first luminous sources emerge~\citep{Tashiro:2005ua, Furlanetto:2006jb, Pritchard:2011xb, Fialkov:2014wka, Kunze:2018cnn}. These scenarios can therefore be probed through their imprint on upcoming 21-cm measurements.

One such possibility is primordial magnetic fields (PMFs), which provide a well-motivated extension of standard cosmology that can be probed through their impact on small-scale structure. Magnetic fields are observed in galaxies and clusters at microGauss ($\mu$G) strengths and may extend into the intergalactic medium at nanoGauss (nG) levels \citep{1994RPPh...57..325K, Clarke:2000bz, Govoni:2004as, Neronov_2010, Taylor_2011, Beck2013}. Although the origin of these large-scale magnetic fields remains uncertain, a compelling possibility is that they were generated during inflation or cosmological phase transitions early in radiation domination~\citep{Grasso:2000wj, Widrow:2002ud, Kandus:2010nw, Durrer:2013pga, Subramanian:2015lua, Vachaspati:2020blt}. This makes PMFs a compelling and testable extension of standard structure-formation models, with 21-cm observations directly probing the allowed excess of small-scale power.

While PMFs on large scales are constrained by existing CMB and large-scale structure measurements~\citep{Kahniashvili:2008hx, Planck:2015zrl, Jedamzik:2018itu, Paoletti:2019pdi, Paoletti:2022gsn}, their impact on smaller, currently unconstrained scales could leave distinctive signatures in the 21-cm signal~\citep{Tashiro:2005ua, Kunze:2014eka, Fialkov:2014wka, 10.1093/pasj/psac015, Venumadhav:2014tqa, Gluscevic:2016gns}. Primordial magnetic fields can seed additional baryonic perturbations, alter the abundance of low-mass halos, and inject energy into the IGM~\citep{Wasserman1978, Gnedin:2000uj, Sethi:2004pe, Tashiro:2005ua, Sethi:2008eq, Yamazaki:2010nf, Schleicher_2010, Pandey_2012, Pandey:2014vga, Fialkov:2014wka}. These effects can modify the timeline of structure formation and EoR and leave detectable imprints on 21-cm observables~\citep{Kunze:2018cnn, 10.1093/pasj/psac015, Schleicher_2011}. They may also help address outstanding problems, such as the origin of cosmic magnetic fields and the photon budget required for reionization~\citep{Miralda-Escude:2003lmt}. By probing scales beyond those accessible to CMB and large scale structure (LSS) measurements, 21-cm observations provide complementary constraints on PMFs.

\begin{figure*}[t]
    \centering
    \includegraphics[width=\textwidth]{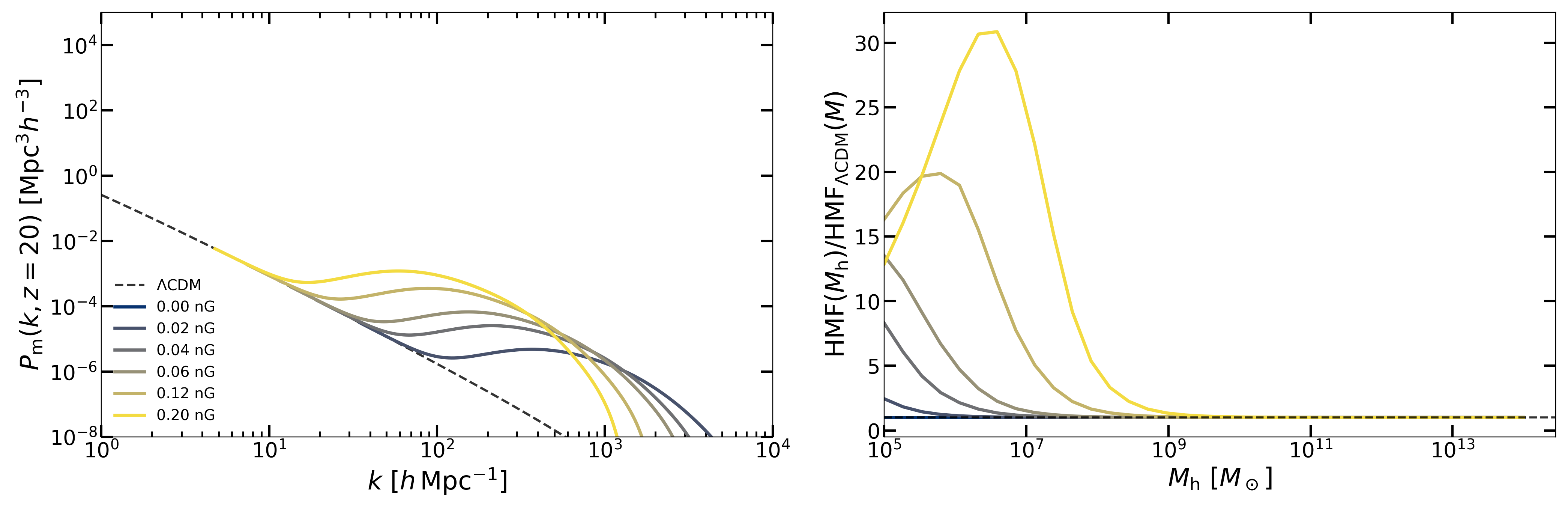}
    \caption{Matter power spectrum (left) and Sheth--Tormen halo mass function ratio relative to $\Lambda$CDM, $\mathrm{HMF}(M_h)/\mathrm{HMF}_{\Lambda\mathrm{CDM}}(M_h)$ (right), at redshift $z=20$ for several primordial magnetic-field (PMF) amplitudes. Increasing PMF strength enhances small-scale power through Lorentz force--sourced perturbations, generating the characteristic high-$k$ bump and damping cutoff in $P_m(k)$. This extra small-scale power boosts the abundance of low-mass halos, producing order-unity to order-of-magnitude enhancements in the halo mass function at $M_h\sim10^5$--$10^8\,M_\odot$, while the ratio approaches unity at high masses where PMF effects are negligible. Although the low-mass HMF can increase strongly, the net SFRD response is more moderate because the star-formation efficiency declines toward low halo masses and the PMF contribution is damped at very high $k$.}
    \label{fig:pk_HMF_main}
\end{figure*}

Interpreting these signatures requires models that balance physical accuracy with computational efficiency. Fully magnetohydrodynamic simulations are prohibitively expensive for parameter-space exploration~\citep{Ryu:2008hi, Dubois:2008mz, Marinacci:2015dja}, and most semi-numerical studies including PMFs have relied on simplified power-spectrum parameterizations to capture their small-scale impact~\citep{Sethi:2004pe, Schleicher_2011, Kunze:2014eka}. This motivates the use of a fast, physically grounded, fully analytical framework such as {\tt\string zeus21}~\citep{Munoz:2023kkg, Cruz:2024fsv}.

Building on the PMF formalism of Refs.~\cite{Cruz:2023rmo,Adi:2023doe} and the {\tt\string zeus21} framework of Refs.~\cite{Munoz:2023kkg,Cruz:2024fsv}, we implement a physically motivated PMF contribution to the linear matter power spectrum within {\tt\string zeus21}, enabling end-to-end 21-cm forecasting. We present exposure-matched HERA and SKA Fisher forecasts (730 $\mathrm{h}$ and 6570 $\mathrm{h}$) in a common parameter basis. We then quantify robustness to external priors, fiducial PMF amplitude, noise-model assumptions, and finite-difference step choices at fixed $n_B=-2.9$ using an expanded set of {\tt\string zeus21} runs. Finally, we introduce a low-dimensional random-forest surrogate to diagnose PMF information content in noiseless {\tt\string zeus21} runs. While the Fisher analysis identifies parameter degeneracies through local sensitivity of the 21-cm signal, the surrogate provides a complementary check that these degeneracies are intrinsic to the computed observables rather than an artifact of the Fisher construction, and is used for interpretive consistency rather than for primary constraints.

This paper is organized as follows. In Section~\ref{sec:pmf_ps}, we summarize PMF physics and the resulting modification to the matter power spectrum. Section~\ref{sec:zeus_implementation} describes how PMFs are incorporated into {\tt\string zeus21}. Section~\ref{sec:fisher_setup} presents the Fisher forecast and the computed instrumental sensitivity. Section~\ref{sec:results} gives the main forecast results, and Section~\ref{sec:discussion_conclusions} discusses implications and concludes.

\section{Primordial Magnetic Fields and the Matter Power Spectrum}
\label{sec:pmf_ps}
In this section, we describe how PMFs generate additional matter fluctuations, proceeding from the magnetic-field power spectrum to the induced matter perturbations and their redshift evolution.  Primordial magnetic fields (PMFs) can modify the growth of cosmic structure by sourcing additional density fluctuations on small scales~\citep{Wasserman1978, Kim:1994zh, Sethi:2004pe, Yamazaki:2010nf}. We model a stochastic, homogeneous, and isotropic PMF with comoving rms strength $\sigma_{B,0}$ (smoothed on $1~\mathrm{Mpc}$ scales) and a power-law spectral index $n_B$~\citep{Subramanian:2015lua, Durrer:2013pga}. The index $n_B$ sets the scale dependence of magnetic fluctuations, with $n_B \to -3$ corresponding to a scale-invariant spectrum on large scales.

Although PMFs are statistically homogeneous and isotropic, their impact on structure formation is confined to a limited range of spatial scales. Prior to recombination, magnetic fields are tightly coupled to the baryon–photon fluid. Radiative viscosity damps Alfv\'en waves and erases magnetic fluctuations below a characteristic scale, removing power on small spatial scales~\citep{PhysRevD.57.3264, PhysRevD.58.083502, Kahniashvili:2009qi}. Following previous works \citep{Cruz:2023rmo, Adi:2023doe}, we capture this suppression using a Gaussian cutoff:
\begin{equation}
P_B(k) = A_B\,k^{n_B}\,\exp\!\left[-2\left(\frac{k}{k_{\rm cut}}\right)^2\right],
\label{eq:PB_damped}
\end{equation}
where $k_{\rm cut}(\sigma_{B,0},n_B)$ sets the effective damping scale for PMF fluctuations. 

These processed magnetic fields source baryonic density perturbations via the Lorentz force. The resulting magnetically induced matter fluctuations are encoded in a mode-coupling term $\Pi(k)$, which captures how pairs of magnetic-field modes combine to generate density perturbations at specific scales $k$. In Fourier space, this corresponds to a convolution over pairs of magnetic modes $\mathbf{k}_1$ and $\mathbf{k}_2$ that form closed triangles with $\mathbf{k} = \mathbf{k}_1 + \mathbf{k}_2$~\citep{Kim:1994zh, Adi:2023doe, Cruz:2023rmo}.
We compute $\Pi(k)$ using an external PMF power-spectrum solver implementing this formalism. Its shape is
\begin{align}
\Pi(k) &= \int_0^\infty k_1^2\,\mathrm{d}k_1 \int_{-1}^{1} \mathrm{d}\mu\,
P_B(k_1)\, P_B\!\left(\sqrt{k^2 + k_1^2 - 2kk_1\mu}\right)
\nonumber \\ &\qquad\times \left[k^2 + \left(k^2 - 2kk_1\mu\right)\mu^2\right],
\label{eq:Pi_full}
\end{align}
where $\mu$ is the cosine of the angle between $\mathbf{k}$ and $\mathbf{k}_1$.

After recombination, magnetic pressure introduces a magnetic Jeans scale that inhibits collapse below a corresponding spatial scale~\citep{Wasserman1978, Kim:1994zh, Sethi:2004pe} for $\Pi(k)$. This post-recombination suppression is incorporated through the same choice of $k_{\rm cut}$ and a high-$k$ taper that guarantees numerical stability without affecting the physically relevant regime (see Section $II.D$ of ~\citep{Cruz:2023rmo}).

The redshift evolution of PMF-induced matter perturbations is described by a magnetic growth factor $M(t)$, obtained from the linearized matter perturbation equation with a Lorentz-force source term~\citep{Cruz:2023rmo, Adi:2023doe}. The fractional matter overdensity can be written as a sourced solution $\delta_m(\mathbf{x},t) \propto M(t)\,v(\mathbf{x})$, where $v(\mathbf{x})$ encodes the spatial structure of the magnetic forcing. The time-dependent factor $M(t)$ therefore satisfies
\begin{equation}
\ddot{M}(t) + 2H(t)\dot{M}(t) - 4\pi G\,\rho_m(t)\,M(t) = \frac{1}{a^3(t)},
\label{eq:M_ODE}
\end{equation}
with initial conditions at recombination $M(t_{\rm rec}) = \dot{M}(t_{\rm rec}) = 0$~\citep{Cruz:2023rmo}. The PMF-induced matter power spectrum is then
\begin{equation}
P_{\rm PMF}(k,z) = M^2(z)\,\Pi(k).
\end{equation}
The total linear matter power spectrum is
\begin{equation}
P_{\mathrm{tot}}(k,z) = P_{\Lambda\mathrm{CDM}}(k,z) + P_{\mathrm{PMF}}(k,z),
\label{eq:totalP}
\end{equation}
where $P_{\Lambda\mathrm{CDM}}(k,z)$ is computed using the CLASS Boltzmann code~\citep{Lesgourgues:2011re, Blas:2011rf, Lesgourgues:2011rg, Lesgourgues:2011rh} and $P_{\mathrm{PMF}}(k,z)$ is provided by the external solver. We assume adiabatic and PMF-induced modes are uncorrelated, so no cross term is included. Figure~\ref{fig:pk_HMF_main} shows how this additional PMF contribution enhances small-scale power at large $k$.

\begin{figure*}[t]
    \centering
    \includegraphics[width=\textwidth]{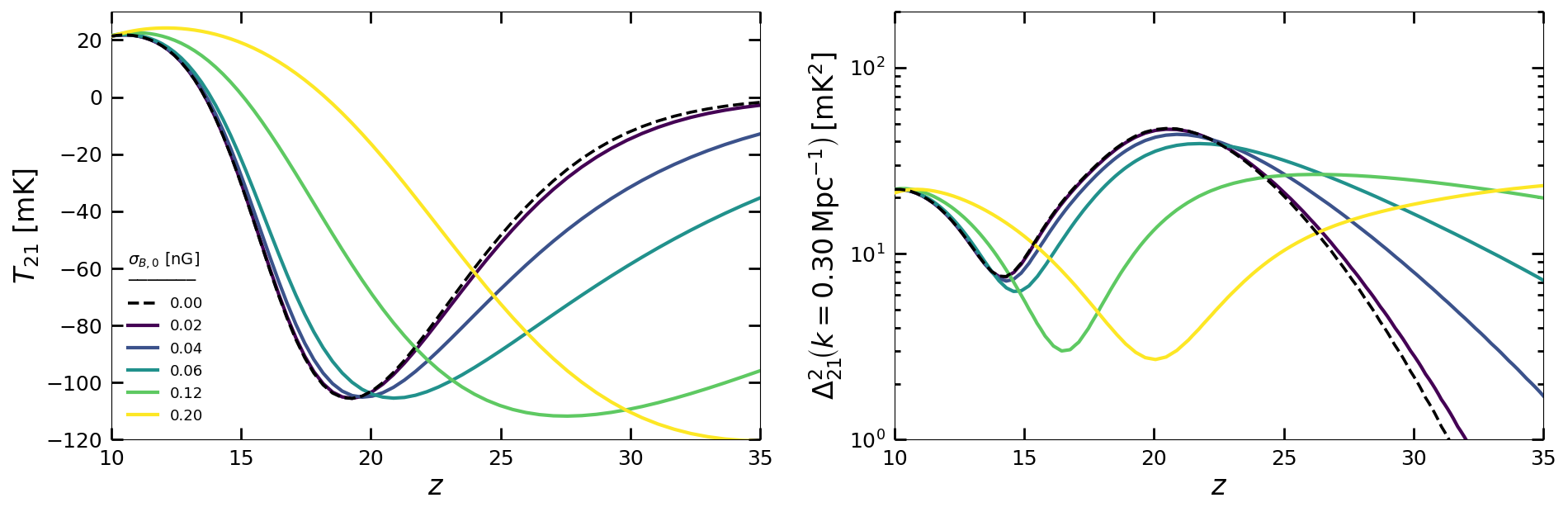}
    \caption{Global and fluctuating 21-cm observables for varying PMF amplitude $\sigma_{B,0}$ at fixed $n_B=-2.9$ (all other parameters held at their fiducial values). \textbf{Left:} sky-averaged brightness temperature $T_{21}(z)$. \textbf{Right:} reduced 21-cm power spectrum $\Delta^2_{21}(k,z)$ evaluated at $k=0.30\,\,\mathrm{Mpc}^{-1}$. Increasing $\sigma_{B,0}$ produces clear, correlated shifts in both the timing and amplitude of the global signal and power-spectrum evolution.}
    \label{fig:t21_delta21} 
\end{figure*}

The additional small-scale power enhances the matter variance $\sigma^2(M,z)$, increasing the abundance of low-mass halos and leading to earlier structure formation. For sufficiently large $\sigma_{B,0}$ and blue spectral indices ($n_B \gtrsim -2.9$), PMFs can dominate $\sigma(M,z)$ at low masses, substantially boosting early halo formation~\citep{Sethi:2004pe, 10.1093/pasj/psac015}. In our implementation, this transition occurs for $\sigma_{B,0}\gtrsim 10^{-2} \mathrm{nG}$, where the PMF-induced contribution overtakes the $\Lambda$CDM variance at $M_h\sim10^5$--$10^8M_\odot$ (Fig.~\ref{fig:pk_HMF_main}).

To propagate these effects into observable predictions, we incorporate $P_{\mathrm{tot}}(k,z)$ directly into {\tt\string zeus21} as a tabulated function in $(\log k, z)$ and evaluated using cubic-spline interpolation, with $P_{\rm PMF}=0$ enforced outside the tabulated range. Once supplied, {\tt\string zeus21} automatically recomputes the mass variance, halo abundance (using the Sheth--Tormen formalism~\citep{Sheth:1999mn}), and collapse fraction, enabling direct comparison to a baseline $\Lambda$CDM scenario without modifying downstream astrophysical modules. The resulting enhancement of small-scale power is shown in the left panel of Fig.~\ref{fig:pk_HMF_main}, with the corresponding increase in low-mass halo abundance shown in the right panel.

\section{21-cm Modeling}
\label{sec:zeus_implementation}
{\tt\string Zeus21} is a fast, fully analytical framework for modeling the global and fluctuating 21-cm signal during Cosmic Dawn and the Epoch of Reionization. It tracks the coupled evolution of density, ionization, and temperature fields, reproducing the nonlinear and nonlocal structure of radiation backgrounds to within $\sim$5\% of established semi-numerical simulations such as \textsc{21cmFAST}~\citep{Munoz:2023kkg, Cruz:2024fsv, Mesinger_2010}. Unlike semi-numerical approaches that rely on repeated real-space filtering, FFT-based bubble finding, or photon Monte Carlo propagation~\citep{Mesinger_2010, Watkinson:2018efd}, {\tt\string zeus21} performs these calculations analytically, yielding a speedup of $\sim10^3$–$10^4$ while maintaining comparable accuracy~\citep{Munoz:2023kkg,Cruz:2024fsv}. This combination of speed and accuracy makes {\tt\string zeus21} particularly well suited for our purposes, as it enables efficient exploration of beyond-$\Lambda$CDM scenarios and exotic physics such as PMF-induced modifications to the matter power spectrum, while remaining computationally tractable across high-dimensional parameter spaces. The code interfaces directly with the \textsc{CLASS} Boltzmann solver to obtain linear matter transfer functions, while the adiabatic gas temperature evolution is computed self-consistently within {\tt\string zeus21}~\citep{Lesgourgues:2011re, Blas:2011rf, Lesgourgues:2011rg, Lesgourgues:2011rh}.

The {\tt\string zeus21} code computes the radiation fields emitted by two separate stellar populations. Pop III stars reside in molecular-cooling minihalos ($M_h \sim 10^6$--$10^8\,M_\odot$), while Pop II stars form in atomic-cooling halos ($M_h \gtrsim 10^8\,M_\odot$)~\citep{Bromm:2013iya}. Each population is assigned its own star-formation efficiency $f_\star(M_h,z)$ and spectral energy distribution. Feedback from Lyman–Werner radiation and baryon–dark matter relative velocities suppress star formation in low-mass halos, introducing scale-dependent structure in the resulting radiation fields~\citep{Fialkov:2014kta, Cruz:2024fsv}. These star-formation histories source ionizing, Lyman-$\alpha$, and X-ray radiation backgrounds. Rather than evolving these backgrounds through explicit radiative-transfer simulations, {\tt\string zeus21} computes them analytically using effective bias expansions and window functions that capture their nonlocal propagation. This approach enables rapid and self-consistent evaluation of the radiation fields across parameter space, making it well suited for exploring scenarios in which PMFs shift the timing and amplitude of early star formation~\citep{Furlanetto:2006jb, Munoz:2023kkg}. The resulting radiation fields are then mapped into the local IGM quantities $x_{\mathrm{HI}}(\mathbf{x},z)$, $T_K(\mathbf{x},z)$, and $x_\alpha(\mathbf{x},z)$ using analytical expressions. These fields determine the spin temperature and therefore the differential 21-cm brightness temperature.

The differential 21-cm brightness temperature can be written as
\begin{equation}
T_{21}(\mathbf{x},z)\propto x_{\mathrm{HI}}(\mathbf{x},z)\,\left(1+\delta_b-\delta_v\right)\left(1-\frac{T_\gamma}{T_S}\right),
\end{equation}
where $T_S$ is the spin temperature, $T_\gamma$ is the CMB temperature, $\delta_b$ is the baryon overdensity, and $\delta_v$ encodes line-of-sight velocity gradients. The spin temperature depends on Ly$\alpha$ coupling, collisional coupling, and radiative processes~\citep{Pritchard:2011xb, Furlanetto:2006jb}.

The corresponding power-spectrum observable used throughout this work is defined as
\begin{equation}
\left\langle \delta T_{21}(\mathbf{k},z)\,\delta T_{21}(\mathbf{k}',z)\right\rangle
=(2\pi)^3\delta_D(\mathbf{k}+\mathbf{k}')\,P_{21}(k,z).
\end{equation}
We work with the dimensionless form $\Delta_{21}^2(k,z) = k^3 P_{21}(k,z)/2\pi^2$ in our analysis.

\begin{figure*}[t]
    \centering
    \includegraphics[width=\textwidth]{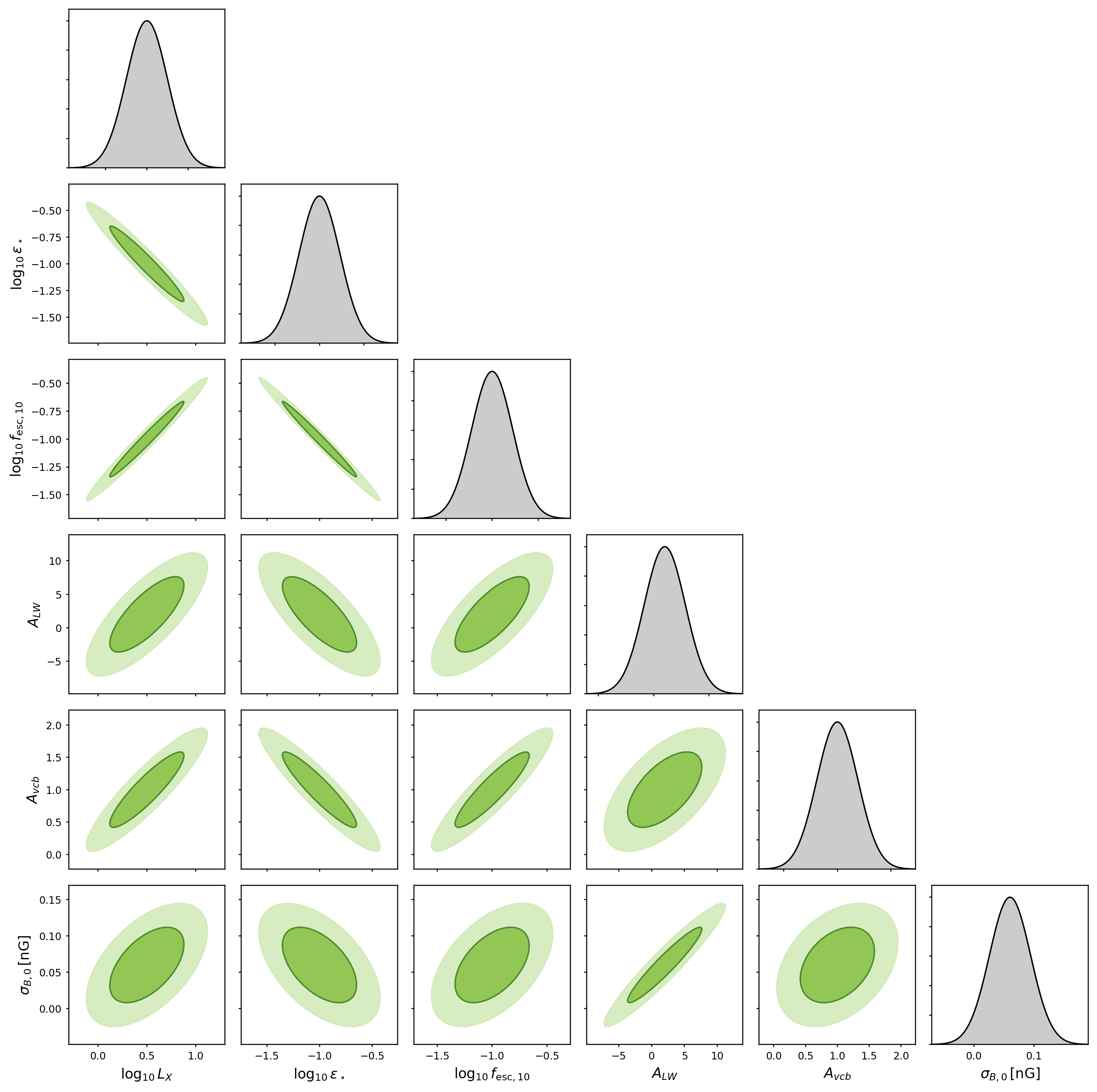}
    \caption{Marginalized Fisher constraints for HERA at fiducial $\sigma_{B,0}=0.06\,\mathrm{nG}$ (fixed $n_B=-2.9$), using the 21-cm + CMB prior configuration (Planck 2018 prior set). This figure shows the long 6570-hour campaign only. In each 2D panel, shaded regions denote the marginalized 68\% (dark green) and 95\% (light green) confidence regions; the dark green line marks the 68\% boundary. Diagonal panels show the corresponding 1D marginalized posteriors. As expected with external CMB priors, the cosmological block $(h,\omega_b,\omega_c,A_s)$ is strongly compressed, while the dominant remaining covariance lies in the astrophysical/PMF sector $(\log_{10}L_X,\log_{10}\epsilon_\star,\log_{10}f_{{\rm esc},10},A_{LW},A_{vcb},\sigma_{B,0})$.}
    \label{fig:fisher_06ng_planckpriors_hera_long}
\end{figure*}

\begin{figure*}[t]
    \centering
    \includegraphics[width=\textwidth]{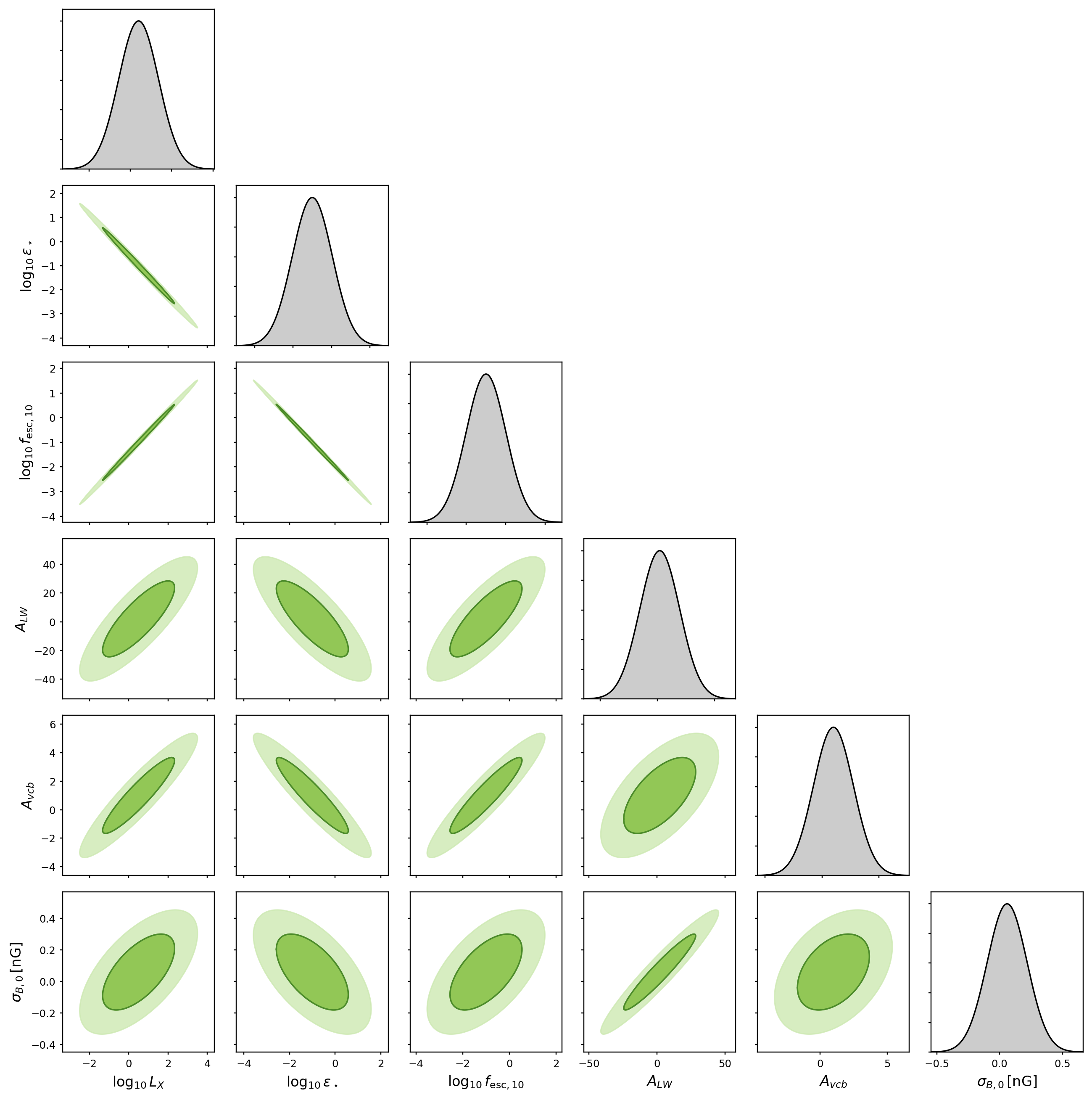}
    \caption{Same setup as Fig.~\ref{fig:fisher_06ng_planckpriors_hera_long}, but for the HERA short campaign (730 h) with 21-cm + CMB priors at fiducial $\sigma_{B,0}=0.06\,\mathrm{nG}$ and fixed $n_B=-2.9$. Relative to the long run, the contours broaden, with the largest degradation in the astrophysical/PMF sector.}
    \label{fig:fisher_06ng_planckpriors_hera_short}
\end{figure*}

\begin{figure*}[t]
    \centering
    \includegraphics[width=\textwidth]{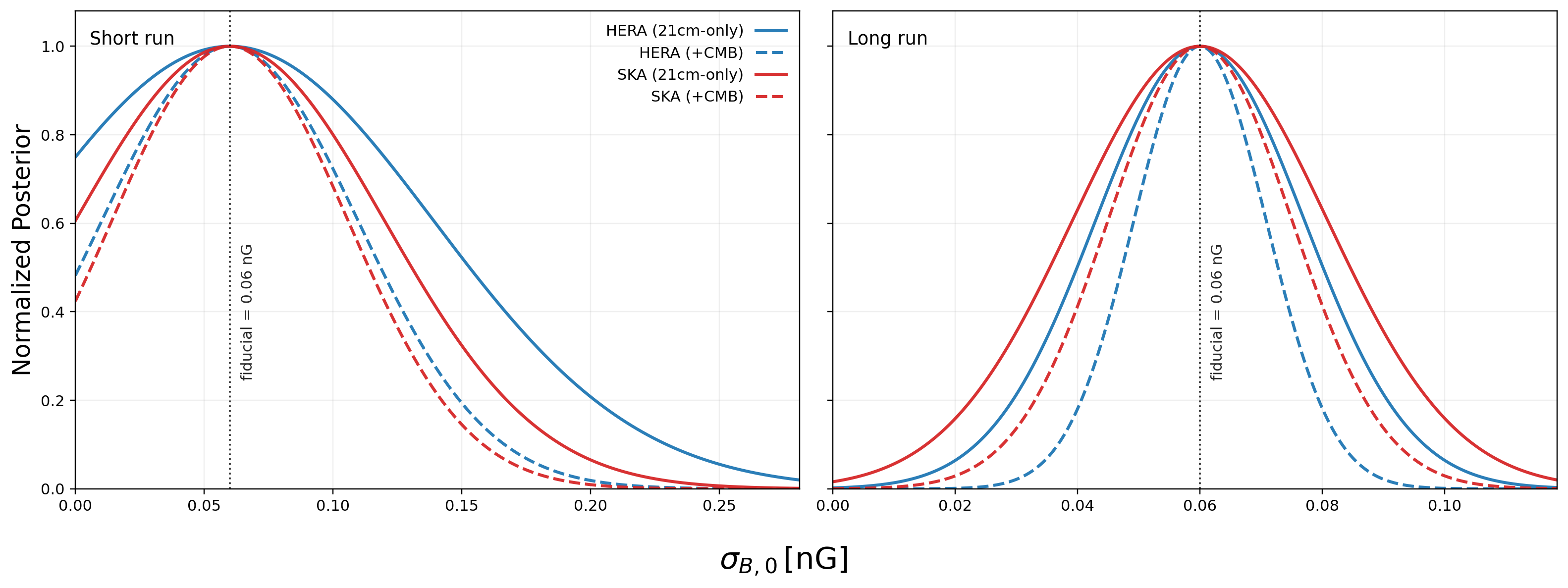}
    \caption{
    One-dimensional marginalized constraints on the PMF amplitude $\sigma_{B,0}$ for fiducial $\sigma_{B,0}=0.06\,\mathrm{nG}$ (vertical dotted line), comparing HERA and SKA with (dashed) and without (solid) CMB priors. The left panel shows the short observing campaign; the right panel shows the long campaign.  In both instruments, adding CMB priors narrows the posterior, with the strongest relative effect for HERA in the long campaign where cosmological degeneracies are most efficiently broken. SKA improves over HERA in the short-campaign regime, while in the long-campaign regime the two experiments become comparable once cosmological information is included, indicating that the remaining uncertainty is increasingly set by shared PMF--astrophysical degeneracy structure rather than raw sensitivity alone.
    }
    \label{fig:oned_sigmaB_hera_ska_06ng}
\end{figure*}

Figure~\ref{fig:t21_delta21} shows the corresponding 21-cm response. As the magnetic amplitude \(\sigma_{B,0}\) increases, the main global-signal features move to higher redshift, and \(\Delta_{21}^2(k,z)\) changes in both timing and amplitude. This behavior arises because PMFs enhance early structure formation, accelerating coupling, heating, and ionization histories, which jointly shift both \(\bar T_{21}(z)\) and $\Delta^2_{21}(k,z)$.

\section{Instrumental Sensitivity Forecasting}
\label{sec:fisher_setup}
We forecast the detectability of PMF-induced signatures in the 21-cm signal for upcoming interferometric experiments, focusing on the Hydrogen Epoch of Reionization Array (HERA) and the Square Kilometre Array (SKA). These instruments probe complementary ranges in wavenumber and redshift, providing sensitivity to the small-scale structure relevant for PMF signatures. For HERA, we adopt a configuration following \citet{Cruz:2023rmo} with the hex-11 array including outriggers, \(8\,\mathrm{MHz}\) bandwidth, \(15\) frequency channels, and \(10.7\,\mathrm{s}\) integrations, assuming moderate foregrounds. For SKA, we use a wide configuration based on the SKA-LOW1 core profile with \(8\,\mathrm{MHz}\) bandwidth, \(60\) channels, \(10.7\,\mathrm{s}\) integrations, and a maximum baseline of \(3\,\mathrm{km}\), again assuming moderate foregrounds. We consider two observational campaigns: a short campaign with \(365\) days at \(2~\mathrm{hr/day}\), and a long campaign with \(1095\) days at \(6~\mathrm{hr/day}\). Thermal and sample noise are computed using {\tt 21cmSense} for these configurations~\citep{Pober:2012zz, Pober:2013jna}. Foreground and wedge exclusions are encoded directly in the noise grids, and masked modes are removed by construction. Unless otherwise noted, forecasts use these baseline noise grids, with alternative noise assumptions explored in Sec.~\ref{sec:results_expanded_campaign}.

To assess whether PMF-induced modifications to the 21-cm signal are observable, we forecast parameter constraints using a Fisher-matrix formalism applied to the 21-cm reduced power spectrum $\Delta^2_{21}(k,z)$.
Our baseline parameter vector is
\begin{equation}
\begin{split}
\{\;h,\omega_b,\omega_c,A_s,\log_{10}L_X,\log_{10}\epsilon_\star,\\
\log_{10}f_{{\rm esc},10},A_{LW},A_{vcb},\sigma_{B,0}\;\}.
\end{split}
\end{equation}
The fiducial values for the non-cosmological parameters, respectively,  are
\[
(0.50,-1.00,-1.00,2.00,1.00,0.06\,\mathrm{nG}).
\]
Cosmological parameters are fixed to the Planck 2018 best-fit values. The Fisher matrix is
\begin{equation}
F_{ij}=\sum_{(k,z)\in\mathcal{M}}
\frac{1}{\sigma^2_{\Delta^2}(k,z)}
\frac{\partial \Delta^2_{21}}{\partial \theta_i}
\frac{\partial \Delta^2_{21}}{\partial \theta_j}
+F^{\rm prior}_{ij},
\label{eq:fisher_def}
\end{equation}
where $\mathcal{M}$ is the set of valid bins and $\sigma_{\Delta^2}(k,z)$ is provided by the {\tt 21cmSense} sensitivity model.
Derivatives are computed via finite differences around a fiducial model,
\begin{equation}
\frac{\partial \Delta^2_{21}}{\partial \theta_i}\approx
\frac{\Delta^2_{21}(\theta_i+\Delta\theta_i)-\Delta^2_{21}(\theta_i-\Delta\theta_i)}
{2\Delta\theta_i},
\end{equation}
using a 5\% fractional step for astrophysical parameters. When varying astrophysical parameters associated with Pop II star formation (e.g., $\epsilon_\star$, $L_X$), we simultaneously vary their Pop III counterparts to maintain a consistent mapping between stellar populations. For the $\sigma_{B,0}=0$ control, we use one-sided derivatives where needed and additionally evaluate the recasted variable in \(\beta_B\equiv\sigma_{B,0}^2\) to improve numerical stability near the null boundary where PMF contributions vanish.

We consider two forecast configurations: a 21-cm only case without external priors, and a 21-cm + CMB case including diagonal Planck 2018 priors on \(h,\omega_b,\omega_c,A_s\)~\citep{Planck:2018vyg}. A diagonal covariance is assumed in $(k,z)$ space,
\[
C_{ab}=\sigma^2_{\Delta^2}(k_a,z_a)\,\delta_{ab},
\]
so inter-bin correlations are neglected. The analysis window spans \(k\in[0.0792,\,0.2964]\,\mathrm{Mpc}^{-1}\) and \(z\in[10,35]\); the effective number of usable modes differs between HERA and SKA due to their distinct baseline distributions and foreground filtering, and is determined internally by the {\tt 21cmSense} sensitivity calculations.

A fixed spectral index \(n_B=-2.9\) is adopted as a near scale-invariant benchmark. Such spectra concentrate power toward smaller scales while remaining consistent with existing CMB upper limits on PMFs, allowing for comparatively larger values of the magnetic-field amplitude \(\sigma_{B,0}\) without violating current constraints. Forecasts are presented for a fiducial value \(\sigma_{B,0}=0.06\,\mathrm{nG}\), and dependence on this choice is assessed by evaluating a grid of {\tt\string zeus21} models spanning a range of \(\sigma_{B,0}\) values.

\begin{figure*}[t]
    \centering
    \includegraphics[width=\textwidth]{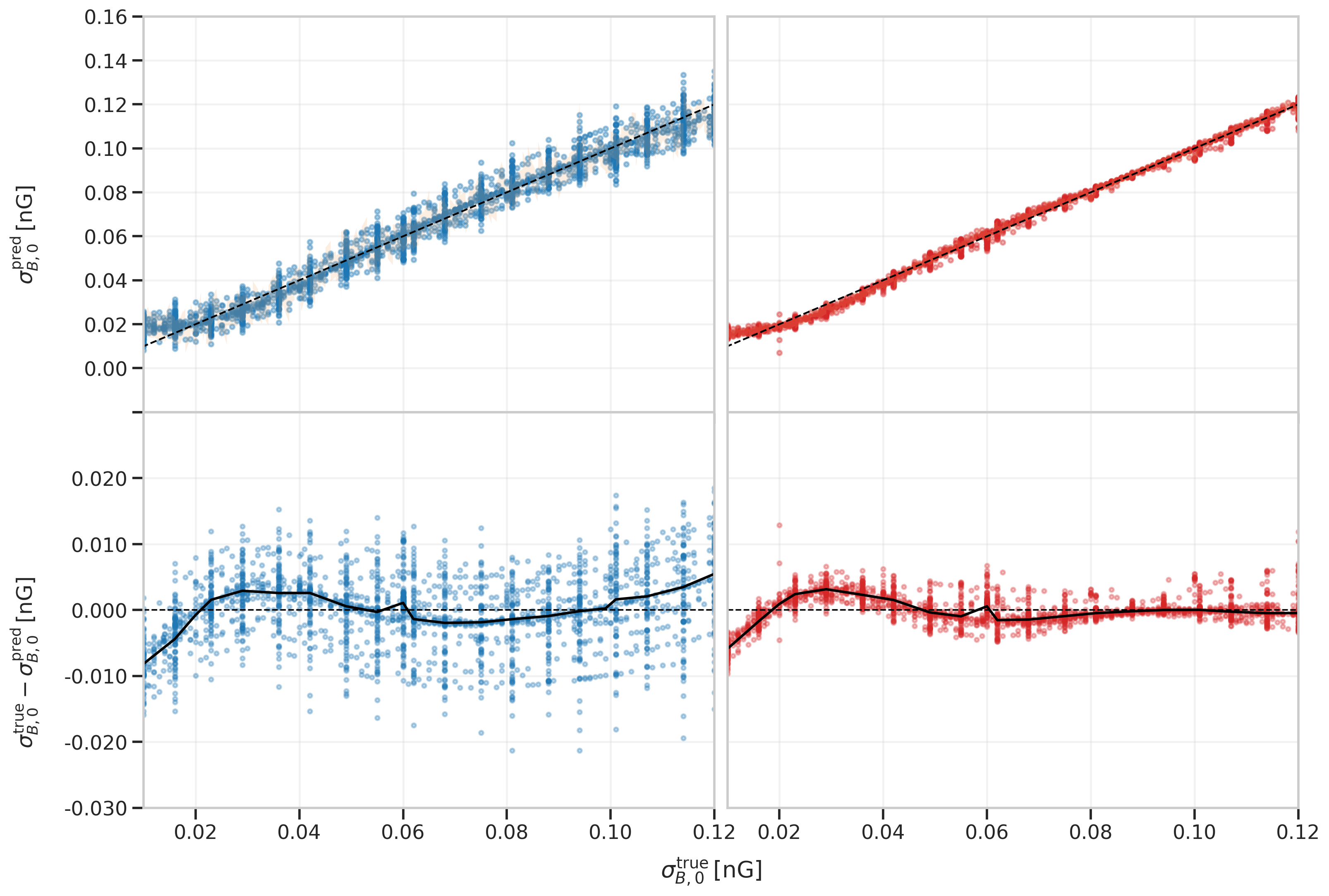}
\caption{Random-forest prediction of the PMF amplitude $\sigma_{B,0}$ from noiseless {\tt\string zeus21} outputs, with each point corresponding to one run. Top panels show predicted versus true $\sigma_{B,0}$ for a 2-feature model (left), using $(T_{21}(z{=}25),\,x_e(z{=}30))$, and a 4-feature model (right), which additionally includes $(A_{LW},A_{vcb})$. The dashed line marks the 1:1 relation. Bottom panels show residuals $(\sigma_{B,0}^{\rm true}-\sigma_{B,0}^{\rm pred})$ versus true $\sigma_{B,0}$, with binned median trends overlaid. The visible clustering at specific $\sigma_{B,0}$ values reflects the discrete sampling of our {\tt\string zeus21} grid and is not a physical feature.
Adding $(A_{LW},A_{vcb})$ is associated with improved test-set performance (2-feature: $R^2=0.9679$, $\mathrm{RMSE}=5.8\times10^{-3}\,\mathrm{nG}$; 4-feature: $R^2=0.9942$, $\mathrm{RMSE}=2.4\times10^{-3}\,\mathrm{nG}$) and reduced residual structure, indicating improved capture of PMF--astrophysical degeneracy directions. We use this RF model only as an alternative diagnostic, with all primary constraints in this work coming from the Fisher analysis.}
    \label{fig:rf_calibrator}
\end{figure*}

\section{Results}
\label{sec:results}
This section examines four aspects of our forecasts: the impact of PMFs on observable 21-cm signals, the parameter degeneracies that shape PMF constraints, the robustness of the HERA--SKA comparison to modeling assumptions and numerical choices, and a surrogate-based internal consistency check. To characterize robustness, we perform more than 14,000 {\tt\string zeus21} evaluations spanning PMF and astrophysical parameter variations, derivative step sizes, and targeted stress-test configurations. We find that PMFs systematically shift 21-cm signals to earlier times, that their constraints are primarily limited by degeneracies with astrophysical parameters, and that relative instrument performance depends sensitively on noise assumptions. We now turn to the resulting parameter constraints and degeneracy structure.

\subsection{Baseline Fisher structure}
\label{sec:results_fisher_structure}
For a fiducial model with $\sigma_{B,0}=0.06\,\mathrm{nG}$ (fixed $n_B=-2.9$), PMF constraints are primarily limited by degeneracies with astrophysical parameters. Figures~\ref{fig:fisher_06ng_planckpriors_hera_long} and~\ref{fig:fisher_06ng_planckpriors_hera_short} show the corresponding HERA Fisher forecasts with CMB priors for the long and short campaigns, respectively. These corner plots are restricted to the astrophysical+PMF block in order to highlight the residual degeneracy structure after cosmological parameters are constrained by the Planck priors. The impact of these priors on the full 10-parameter constraints is shown in Appendix~\ref{app:fisher_table}, where Table~\ref{tab:fisher_constraints_21cm_only} presents the 21-cm-only case and Table~\ref{tab:fisher_constraints_cmb} includes CMB priors.

Within this sector, the dominant degeneracy involves $\sigma_{B,0}$ and the Lyman--Werner feedback parameter $A_{\rm LW}$. This direction is clearly positive in both the short and long campaign forecasts. Additional, weaker degeneracies are present between $\sigma_{B,0}$ and parameters controlling X-ray heating and star formation efficiency, $\log_{10}L_X$ and $\log_{10}\epsilon_\star$. The covariance with $\log_{10}f_{{\rm esc},10}$ is present but subdominant, and this parameter is primarily weakly constrained, especially in the short campaign. These trends arise because both PMFs and astrophysical parameters control the timing and amplitude of early radiation backgrounds. Increasing $A_{\rm LW}$ suppresses star formation in low-mass halos, which can be offset by increasing $\sigma_{B,0}$ to restore early structure formation, producing the observed positive degeneracy. A similar, though weaker, effect appears for $\log_{10}L_X$ where higher X-ray efficiency shifts the heating history earlier, which can be partially mimicked by larger $\sigma_{B,0}$ and produce a positive correlation. In contrast, the degeneracy with $\log_{10}\epsilon_\star$ is negative, indicating that increased star formation efficiency can be partially compensated by a lower PMF amplitude. Some parameters remain only weakly constrained even after including CMB priors. In particular, the escape fraction $f_{{\rm esc},10}$ exhibits broad, weakly bounded contours spanning multiple orders of magnitude. This reflects its limited empirical constraints during Cosmic Dawn, where direct measurements of ionizing escape are not available~\citep{Ciardi:2003ia, Gnedin:2007pw, Kuhlen:2012vy, Wise:2014vwa}.

In addition, parts of the allowed parameter space extend into astrophysically implausible or unphysical regimes. Negative values of $A_{vcb}$ correspond to baryon--dark matter relative velocities enhancing, rather than suppressing, star formation, which is not physically motivated. The upper range of the $\log_{10}L_X$ contours at the $2\sigma$ level may also be in tension with constraints from X-ray binary population models and high-redshift observations (e.g., \citealt{Fragos:2012vf,Fragos:2013bfa}). These features reflect the fact that the Fisher analysis is driven by degeneracy structure in the 21-cm signal alone, without enforcing external astrophysical priors. As expected, the long campaign reduces the extent of these allowed regions and narrows the same degeneracy directions relative to the short campaign, without changing their qualitative orientation.

\subsection{HERA--SKA comparison at the baseline}
\label{sec:results_instrument_compare}
For the fiducial ($\sigma_{B,0}=0.06 \mathrm{nG}$), SKA outperforms HERA in short surveys, while the two reach near-parity in long integrations as seen in Fig.~\ref{fig:oned_sigmaB_hera_ska_06ng}. CMB priors tighten constraints for both instruments. We compare matched exposures (730 h short, 6570 h long) using the adopted baseline noise assumptions and instrument support masks. The ranking reflects both support and noise level: SKA contributes broader $(k,z)$ support over the analysis window, while HERA is competitive on a narrower support and can be lower-noise in parts of the long redshift range. Appendix Fig.~\ref{fig:noise_amp_z189} helps interpret this tradeoff by showing the baseline scale-dependent noise contrast at fixed redshift. Full 10-parameter marginalized $1\sigma$ constraints for both prior configurations are provided in Appendix Tables~\ref{tab:fisher_constraints_21cm_only} and~\ref{tab:fisher_constraints_cmb}.

\subsection{Expanded {\tt\string Zeus21} Campaign: Robustness Checks}
\label{sec:results_expanded_campaign}
We assess the sensitivity of our results to noise-model assumptions, derivative step size, and behavior near the null boundary $\sigma_{B,0}=0$. In particular, we test how changes in foreground treatment and instrument configuration affect the accessible Fourier-space region and the inferred signal sensitivity.

To test noise assumptions, we construct three variants using separate {\tt 21cmSense} runs. The baseline case adopts the moderate-foreground configuration described in Sec.~\ref{sec:fisher_setup}. The optimistic case uses the same instrument configurations but switches to an optimistic foreground model, increasing the accessible $(k,z)$ support through less aggressive foreground filtering and reducing noise on the retained modes. For the support-stress case, we deliberately reduce the effective $(k,z)$ coverage of each instrument through configuration changes applied within {\tt 21cmSense}. For HERA, the number of frequency channels is increased from 15 to 30 (with outriggers unchanged), which reduces per-channel sensitivity and shrinks the usable mode set. For SKA, the number of channels is reduced from 60 to 15 and the maximum baseline is reduced from 3000\,m to 1000\,m, directly limiting its accessible $k$-range. These changes alter the Fourier-space coverage through the {\tt 21cmSense} calculation itself rather than through any post-processing mask.

Across these variants, SKA continues to show stronger projected sensitivity in short surveys, as reflected by the narrower marginalized $\sigma_{B,0}$ posterior in Fig.~\ref{fig:oned_sigmaB_hera_ska_06ng}. In long integrations, the relative performance becomes more comparable, with both instruments yielding similar posterior widths once CMB priors are included. Here, sensitivity is defined by the width of the marginalized $\sigma_{B,0}$ constraint. HERA achieves lower noise over a limited mid-redshift interval, which can improve constraints when those redshift ranges dominate the signal contribution. SKA, in contrast, maintains lower noise across a broader redshift range and benefits from wider $(k,z)$ coverage.

For the Fisher matrix, we compute numerical derivatives of the 21-cm power spectrum $\Delta^2_{21}(k,z)$ with respect to each model parameter using finite differences around the fiducial. To test the stability of these derivatives, we scan over multiple fiducial parameter choices and vary the fractional step size used in the finite-difference calculation. Very small step sizes lead to subtraction noise; the derivative is estimated from the difference of two nearly identical model evaluations, making the result sensitive to numerical precision and interpolation error. Larger step sizes reduce this effect but increasingly probe nonlinear response away from the fiducial, biasing the derivative estimate. Across the tested parameter space, a 5\% fractional step in astrophysical parameters provides stable derivatives and is adopted as a practical baseline choice but not claimed as a universal optimum. Near $\sigma_{B,0}=0$, one-sided derivatives and local nonlinearity make upper-limit estimates more step-size dependent. Reparameterizing to $\beta_B\equiv\sigma_{B,0}^2$ improves conditioning, but null-run bounds remain method-dependent within the tested numerical setup and therefore should not be interpreted as a robust global upper limit.

\subsection{Random-forest surrogate check}
\label{sec:results_rf}
We perform an additional internal consistency check using a random-forest regressor trained on noiseless {\tt\string zeus21} outputs. A random forest is a nonparametric machine-learning model that maps a set of input features to a target quantity, here the PMF amplitude $\sigma_{B,0}$. The goal is not to obtain new forecasts, but to test how much PMF information is already present in a compact set of {\tt\string zeus21} outputs and whether the same PMF--astrophysical degeneracy directions seen in the Fisher analysis also appear in a purely data-driven mapping. This exercise neglects instrumental noise and survey effects and is therefore not an observational forecast; instead, it isolates how PMF information is encoded in the modeled signal itself.

The model is trained on a set of {\tt\string zeus21} outputs with finite values of $\sigma_{B,0}$, $T_{21}(z{=}25)$, $x_e(z{=}30)$, $A_{LW}$, $A_{vcb}$, $\log_{10}\epsilon_\star$, and $\log_{10}L_X$. From the full set of astrophysical parameters, these features were selected from a larger set of {\tt\string zeus21} outputs based on their ability to recover $\sigma_{B,0}$ with low error in preliminary tests. This choice isolates a minimal set of observables and controls that capture the dominant PMF signal and its leading degeneracies.

We first train a 2-feature model using only $(T_{21}(z{=}25),\,x_e(z{=}30))$. This gives $R^2=0.9679$ and RMSE $=5.8\times10^{-3}\,\mathrm{nG}$. We then add the astrophysical controls $(A_{LW},A_{vcb})$, giving a 4-feature model with $R^2=0.9942$ and RMSE $=2.4\times10^{-3}\,\mathrm{nG}$, an RMSE reduction of about $58\%$. In Fig.~\ref{fig:rf_calibrator}, the 2-feature model shows a clear residual trend with true $\sigma_{B,0}$, indicating a systematic bias in the mapping. The residuals exhibit coherent curvature as a function of $\sigma_{B,0}$, rather than random scatter, which signals that the model is missing degrees of freedom needed to disentangle PMF amplitude from astrophysical effects. After adding $(A_{LW},A_{vcb})$, the residuals become flatter and more tightly clustered around zero, indicating that these additional variables capture part of the degeneracy structure and reduce the systematic bias in the prediction.

The results show that even low-dimensional summaries of the {\tt\string zeus21} outputs retain substantial information about $\sigma_{B,0}$, and that the same PMF--astrophysical degeneracy structure identified in the Fisher analysis appears in a fully data-driven mapping. This provides an independent validation of the physical interpretation of parameter covariances and highlights the extent to which PMF information is already present in compact observables.

\section{Discussion and Conclusions}
\label{sec:discussion_conclusions}
At fixed $n_B=-2.9$, PMFs produce a coherent and observable signature across structure formation and 21-cm observables. In our {\tt\string zeus21} model, PMF-enhanced small-scale power increases the abundance of early low-mass halos and drives modifications in both the global signal and the 21-cm power spectrum, consistent with Refs.~\citep{Cruz:2023rmo, Sethi:2004pe, 10.1093/pasj/psac015}. We combine these predictions with a Fisher analysis and {\tt 21cmSense} experimental sensitivities to assess the detectability of PMF signatures and their constraining power.

When forecasting instrumental sensitivities under short one-year and long three-year observational runs, a key result is that SKA is favored in short surveys, while long-survey performance depends on noise assumptions. Under the baseline noise model with CMB priors and fiducial $\sigma_{B,0}=0.06\,\mathrm{nG}$, we find $\sigma(\sigma_{B,0})\sim 0.159\,\mathrm{nG}$ (HERA short) versus $\sim 0.139\,\mathrm{nG}$ (SKA short), and $\sim 0.034\,\mathrm{nG}$ (HERA long) versus $\sim 0.033\,\mathrm{nG}$ (SKA long). Across the full range of noise assumptions, short observing runs consistently favor SKA sensitivity, while long-run ordering depends on the assumed noise model. In our baseline and support-stress configurations, HERA is comparable or better over limited redshift intervals that carry substantial Fisher weight. Under optimistic foreground assumptions, SKA retains an advantage through broader retained support and lower effective noise over much of the window.

Planck 2018 priors on $(h,\omega_b,\omega_c,A_s)$ substantially compress cosmological covariance, so the remaining PMF uncertainty is set primarily by PMF--astrophysical degeneracies. The strongest residual directions involve $\sigma_{B,0}$ with source/feedback parameters, especially the $(\log_{10}L_X,\log_{10}\epsilon_\star,\log_{10}f_{{\rm esc},10})$ sector and $A_{\rm LW}$. The degeneracy with $A_{\rm LW}$ is positive, as stronger Lyman--Werner feedback suppresses early star formation and can be offset by increasing $\sigma_{B,0}$. A similar positive correlation appears with $\log_{10}L_X$, while the degeneracy with $\log_{10}\epsilon_\star$ is negative, reflecting the fact that increased star formation efficiency can be partially compensated by a lower PMF amplitude. Overall, this implies that PMFs at the level considered here leave a measurable imprint in 21-cm observables, but their constraints are ultimately limited by degeneracies with astrophysical processes.

At fixed $n_B=-2.9$, our local Fisher uncertainties are broadly consistent with the sensitivity scale shown in Ref.~\citep{Cruz:2023rmo} (their Fig.~7). The comparison is not one-to-one. Their figure is HERA-focused and overlays several ingredients, including HERA Fisher forecasts and external PMF limits from independent probes. Our analysis isolates Fisher constraints within a single, internally consistent inference pipeline and extends the comparison to matched HERA/SKA observing strategies. Even with these differences, our long-run constraints fall in the same high-sensitivity regime. As the previous work concluded, 21-cm measurements may provide competitive, and in this parameter range often stronger, PMF reach than many existing non-21-cm bounds.

As an additional robustness check, we varied both the fiducial point and the finite-difference step sizes used in the Fisher construction. Across the central fiducial range used for our headline forecasts, the inferred $\sigma(\sigma_{B,0})$ remains stable and preserves the main ranking trends. The largest deviations appear near the null boundary ($\sigma_{B,0}=0$) and at the edges of the tested range, where one-sided derivatives and nonlinear responses of $\Delta^2_{21}(k,z)$ lead to less stable curvature estimates. These trends are explicitly validated using the full {\tt\string zeus21} campaign (more than 14,000 evaluations), which samples variations in $\sigma_{B,0}$, astrophysical parameters, and finite-difference step sizes. Across this ensemble, we find that forecasted uncertainties are robust within the central parameter region but become increasingly sensitive to numerical choices near the null boundary. We therefore interpret the quoted PMF uncertainties as local forecasts around stated fiducials, rather than as a global exclusion curve. A 5\% fractional step for astrophysical parameters provides a stable choice over the tested range, although it is not a universal optimum.

We do not include an external $\tau_{\rm CMB}$ prior in the baseline forecasts. While $\tau$ can be computed in {\tt\string zeus21} as a global quantity from the ionization history (and does not require low-redshift power spectra), our Fisher analysis is constructed solely from 21-cm power-spectrum observables. In the current implementation, these observables are evaluated only down to $z \sim 10$. Including a $\tau_{\rm CMB}$ prior would introduce information that is not derived consistently within the same inference pipeline. We thus restrict the baseline forecasts to power-spectrum observables and leave a fully self-consistent incorporation of $\tau_{\rm CMB}$ to future work.

Overall, our results indicate that PMFs at the levels considered here imprint a coherent, potentially detectable signature on both structure formation and 21-cm observables in upcoming surveys under realistic noise assumptions. The dominant limitation is not raw instrumental sensitivity, but degeneracies with astrophysical parameters governing early star formation and feedback. In this way, Cosmic Dawn offers more than a record of the first stars: it may also preserve the small-scale imprint of magnetic fields generated long before the first galaxies formed.

\section*{Data Availability}

The calculations use the methodology, assumptions, fiducial parameter choices, and cited public software described in the article; the project-specific analysis scripts have not been prepared for public release.

\begin{acknowledgments}
K.W. thanks the MD Space Grant Consortium Graduate and Observatory Fellowships for funding this work. K.W. also thanks the LSST-DA Data Science Fellowship Program, which is funded by LSST-DA, the Brinson Foundation, and the Moore Foundation; his participation in the program has benefited this work. H.A.C. was supported by a grant from the Simons Foundation under Grant No. SFI-MPS-SFJ-00010459 and by the National Science Foundation Graduate Research Fellowship under Grant No. DGE2139757.  MK was supported by NSF Grant No.\ 2412361, NASA ATP Grant No.\ 80NSSC24K1226, and the Templeton Foundation.
\end{acknowledgments}

\bibliography{refs}

\clearpage
\onecolumngrid
\appendix

\section{Full Fisher Constraint Tables}
\label{app:fisher_table}
\suppressfloats[t] 
This appendix collects the full marginalized $1\sigma$ constraints for the baseline fiducial setup used in the main text.  Table~\ref{tab:fisher_constraints_21cm_only} shows the 21-cm-only case, and Table~\ref{tab:fisher_constraints_cmb} shows the 21-cm + CMB-prior case. Fiducial parameter values are listed in Section~\ref{sec:fisher_setup}; here we report only forecasted uncertainties and a compact HERA/SKA comparison at fixed long exposure.

\begin{table}[!htbp]
\centering
\caption{Marginalized $1\sigma$ constraints at $\sigma_{B,0}^{\rm fid}=0.06\,\mathrm{nG}$ for the 21-cm-only Fisher analysis. Columns list forecasted uncertainties for HERA and SKA at matched short (730h) and long (6570h) exposures. The final column gives the long-exposure ratio $\sigma_{\rm HERA}/\sigma_{\rm SKA}$ at 6570h; values $>1$ indicate stronger SKA constraints, while values $<1$ indicate tighter HERA constraints.}
\label{tab:fisher_constraints_21cm_only}
\small
\setlength{\tabcolsep}{5pt}
\begin{tabular}{lrrrrr}
\toprule
Parameter & HERA 730h & HERA 6570h & SKA 730h & SKA 6570h & HERA/SKA (6570h) \\
\midrule
h & 2.85 & 0.636 & 2.11 & 0.493 & 1.29 \\
$\omega_b$ & 0.00366 & 0.000887 & 0.00345 & 0.001 & 0.882 \\
$\omega_c$ & 0.0278 & 0.00621 & 0.0207 & 0.00502 & 1.24 \\
$A_s\,(10^{-9})$ & 0.779 & 0.177 & 0.623 & 0.164 & 1.08 \\
$\log_{10}L_X$ & 1.35 & 0.274 & 1.03 & 0.221 & 1.24 \\
$\log_{10}\epsilon_\star$ & 2.11 & 0.453 & 1.67 & 0.384 & 1.18 \\
$\log_{10}f_{{\rm esc},10}$ & 1.39 & 0.293 & 1.152 & 0.264 & 1.11 \\
$A_{LW}$ & 30.7 & 6.40 & 22.3 & 4.80 & 1.33 \\
$A_{vcb}$ & 3.51 & 0.764 & 2.84 & 0.661 & 1.16 \\
$\sigma_{B,0}\,[\mathrm{nG}]$ & 0.257 & 0.0532 & 0.205 & 0.0443 & 1.20 \\
\bottomrule
\end{tabular}

\end{table}

\begin{table}[!htbp]
\centering
\caption{Same as Table~\ref{tab:fisher_constraints_21cm_only}, but including external CMB priors in the Fisher matrix.}
\label{tab:fisher_constraints_cmb}
\small
\setlength{\tabcolsep}{5pt}
\begin{tabular}{lrrrrr}
\toprule
Parameter & HERA 730h & HERA 6570h & SKA 730h & SKA 6570h & HERA/SKA (6570h) \\
\midrule
h & 0.0053999 & 0.0053994 & 0.00540 & 0.005399 & 1.00 \\
$\omega_b$ & 0.0001495 & 0.0001439 & 0.0001496 & 0.00014536 & 0.99 \\
$\omega_c$ & 0.001197 & 0.001146 & 0.001195 & 0.001134 & 1.01 \\
$A_s\,(10^{-9})$ & 0.0497 & 0.0458 & 0.0496 & 0.0453 & 1.01 \\
$\log_{10}L_X$ & 1.21 & 0.250 & 0.838 & 0.186 & 1.35 \\
$\log_{10}\epsilon_\star$ & 1.04 & 0.232 & 0.750 & 0.189 & 1.23 \\
$\log_{10}f_{{\rm esc},10}$ & 1.02 & 0.223 & 0.729 & 0.179 & 1.25 \\
$A_{LW}$ & 17.5 & 3.71 & 11.8 & 2.70 & 1.37 \\
$A_{vcb}$ & 1.76 & 0.384 & 1.43 & 0.348 & 1.10 \\
$\sigma_{B,0}\,[\mathrm{nG}]$ & 0.159 & 0.0343 & 0.139 & 0.0329 & 1.04 \\
\bottomrule
\end{tabular}

\end{table}

\clearpage
\FloatBarrier
\section{Noise-model diagnostic details}
\label{app:noise_diagnostics}
Here, we provide additional context to help interpret the HERA/SKA comparison in the main Fisher results. The first figure shows the underlying noise amplitudes and $k$-coverage at a representative redshift, and the second shows how that noise propagates into the PMF-amplitude Fisher weight as a function of redshift.

In our expanded {\tt\string zeus21} runs, we tested baseline, optimistic, and support-stress noise variants using separate \texttt{21cmSense} runs. The baseline case is the one used for the figures here and for the main quoted constraints. The optimistic and support-stress variants are discussed in Sec.~\ref{sec:results_expanded_campaign}; they change the same basic ingredients (foreground treatment and effective mode support) so are helpful in checking how stable the instrument ranking is.

\begin{figure}[!htbp]
  \centering
  \includegraphics[width=0.7\textwidth]{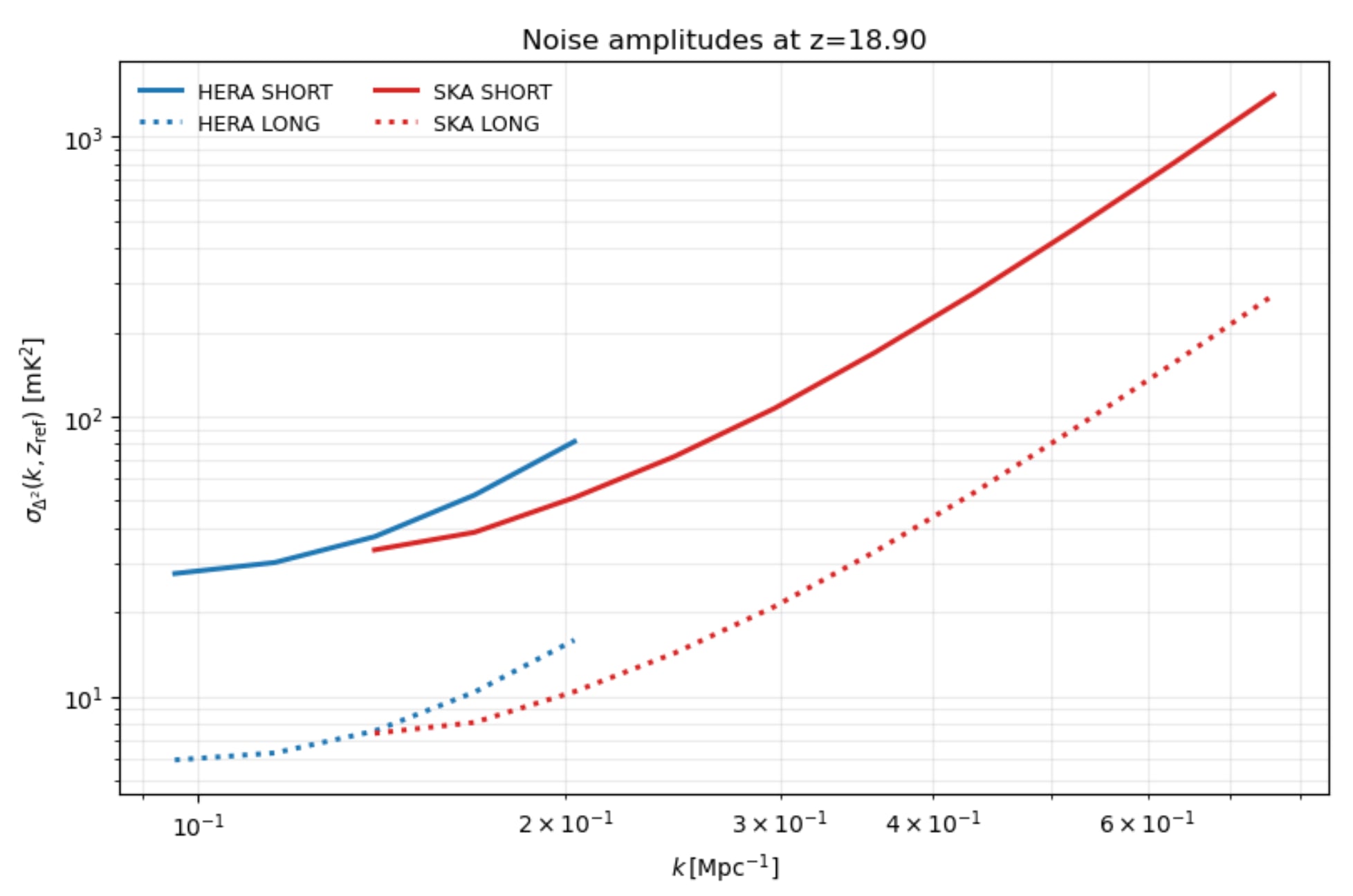}
  \caption{Noise amplitude comparison at fixed redshift $z=18.90$ for matched short/long campaigns. Solid curves denote short (730 h) and dotted curves denote long (6570 h), for HERA and SKA. This panel isolates the scale dependence of $\sigma_{\Delta^2}(k,z_{\rm ref})$ under the baseline noise setup used in the Fisher forecasts.}
  \label{fig:noise_amp_z189}
\end{figure}

\begin{figure}[!htbp]
  \centering
  \includegraphics[width=0.7\textwidth]{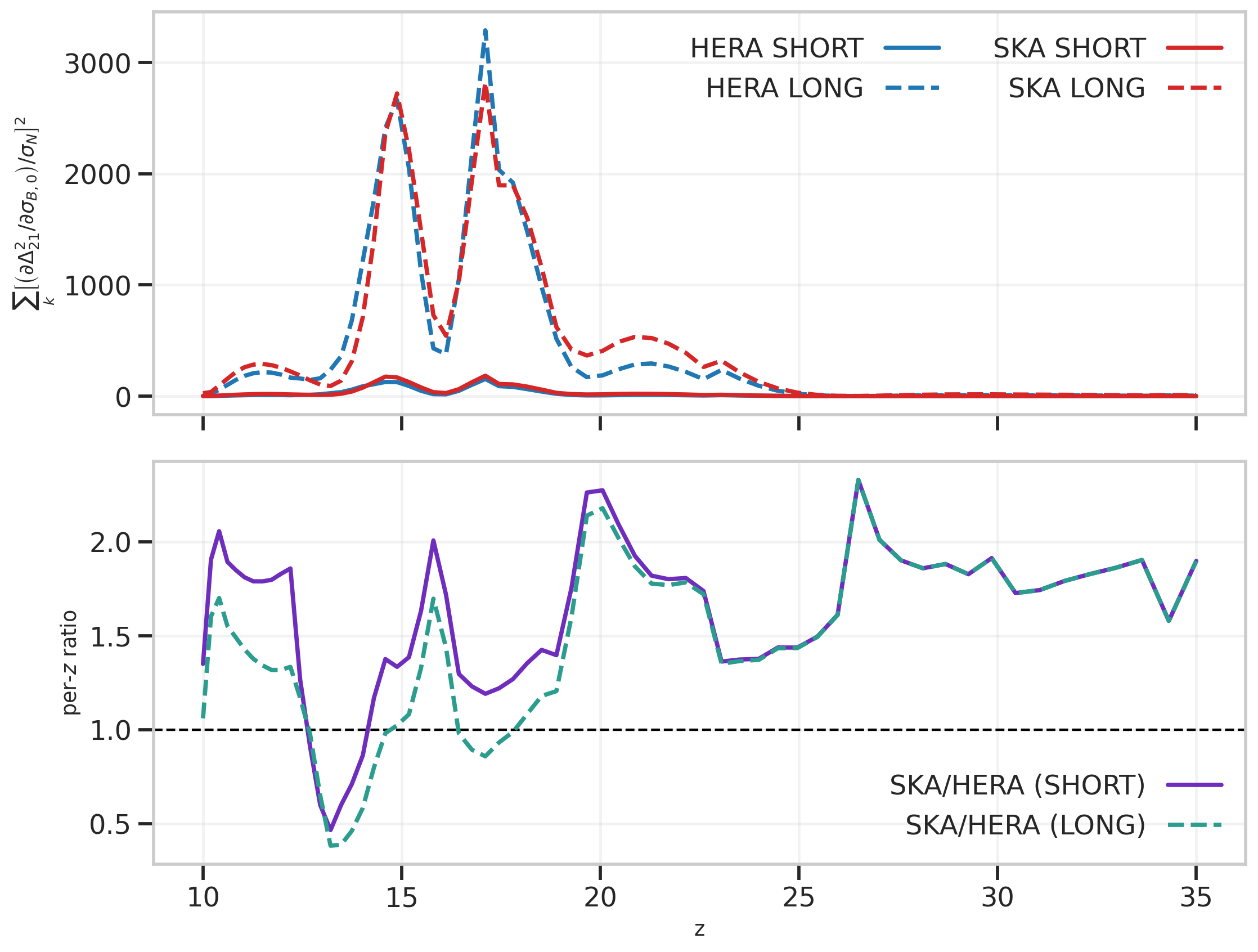}
  \caption{Redshift-resolved Fisher weighting for the PMF-amplitude direction in the baseline setup. \textbf{Top}: per-redshift contribution to the $\sigma_{B,0}$ diagonal term of Eq.~\eqref{eq:fisher_def}, $W_{\sigma_B}(z)=\sum_k[(\partial \Delta^2_{21}/\partial \sigma_{B,0})/\sigma_N]^2$, for HERA and SKA in short (solid) and long (dashed) campaigns. \textbf{Bottom}: SKA/HERA ratio of the same quantity for short and long cases; values above unity indicate larger SKA weight at that redshift, while values below unity indicate larger HERA weight. The dominant weighting lies at intermediate redshifts, with long-campaign differences concentrated in a limited subset of bins.}
  \label{fig:fisher_weight_z}
\end{figure}

Figure~\ref{fig:noise_amp_z189} shows the baseline noise amplitudes at fixed $z=18.90$ for matched short/long observing runs. HERA exhibits stronger coverage at low $k$, while SKA extends to higher $k$ with broader support. Within their overlapping $k$-range, SKA generally achieves lower noise levels. These noise profiles clarify both the accessible regions of $k$-space and where each instrument retains the lowest-noise modes.

Figure~\ref{fig:fisher_weight_z} shows the redshift-resolved contribution to the $\theta_i=\theta_j=\sigma_{B,0}$ diagonal Fisher weight, $W_{\sigma_B}(z)=\sum_k[(\partial\Delta^2_{21}/\partial\sigma_{B,0})/\sigma_N]^2$, evaluated over the same Fisher window and valid support. This is the quantity that directly explains why relative performance can differ between short and long campaigns: the ranking is set by the bins that carry most of the PMF information, not by a single global noise ratio.

\end{document}